\documentclass[letterpaper, paper,11pt]{AAS}	

\usepackage{bm}
\usepackage{amsmath}
\usepackage{gensymb}
\usepackage{subfigure}
\usepackage{flafter}
\usepackage{placeins}
\usepackage{longtable,tabularx}
\usepackage[colorlinks=true, pdfstartview=FitV, linkcolor=black, citecolor= black, urlcolor= black]{hyperref}
\usepackage{overcite}
\usepackage{footnpag}

\PaperNumber{19-681}

\begin{document}

\title{The Long-Term Forecast of Station View Periods for Elliptical Orbits\protect\footnote{\textcopyright\;\MakeLowercase{2019 copyright. all rights reserved.}}}

\author{Andrew J. Graven\thanks{Undergraduate, Department of Mathematics, College of Arts and Sciences, Cornell University, Ithaca, NY, 14853, United States}, Martin W. Lo \thanks{Technologist, Mission Design and Navigation Section, Jet Propulsion Laboratory, California Institute of Technology, Pasadena, CA, 91109, United States}}

\maketitle{}

\begin{abstract}
In a previous paper, using ergodic theory, Lo \cite{Lo1994} derived a simple definite integral that provided an estimate of the view periods of ground stations to satellites. This assumes the satellites are in circular orbits with non-repeating ground tracks under linear $J_2$ perturbations. The novel feature is that this is done without the propagation of the trajectory by employing ergodic theory. This accelerated the telecommunications mission design and analysis by several orders of magnitude and greatly simplified the process. In this paper, we extend the view period integral to elliptical orbits.
\end{abstract}

\section{Introduction}
When planning the placement of ground stations and the trajectories of the satellites communicating with them, it's often important to know how frequently and for what duration the two will be able to directly communicate. This problem amounts to answering the question: for what ratio of total flight time can we expect the satellite to have line of sight communication with it's ground station? Both in the short and long term, the standard method of calculating this statistic is propagating the satellite trajectory over a number of orbits, and averaging the visibility time. For shorter term missions, this sort of trajectory propagation is feasible. However, if the determination of longer term trends is needed, this technique of trajectory propagation becomes impractical due to excessive run time. For example, running the trajectory propagation of a single satellite for 5 integration years on Matlab's ode45 takes $\approx 20$ minutes. When this propagation is needed for a large number of potential satellite trajectories and ground station placements, say fifty potential satellite trajectories and ten potential ground station locations, the required compute time is about a week.

An alternative way to determine this average satellite-ground station visibility ratio is through the use of dynamical systems and statistical mechanics. A key aspect of statistical mechanics is the determination of the long term statistical behavior of a dynamical system, in this case, satellite trajectories. Using this approach, Lo\cite{Lo1994} was able to derive a simple definite integral which yields the expected satellite-ground station visibility ratio to a high degree of speed and accuracy. However, this came with the caveat that the formula is only valid for non-periodic, circular, orbits. Employing a similar technique to that originally used by Dr. Lo, the satellite-ground station visibility formula has been extended to the much wider class of non-periodic elliptical orbits.

Each of the mathematical symbols used throughout the paper are explained in-line and collected in the Notation section at the end of the paper.

\section[The Long Term Ground Station View Period Ratio, p]{The Long Term Ground Station View Period Ratio, $\boldsymbol\rho$}
In both the circular and elliptical cases, the integral yields an approximation of the long-term station view period ratio, $\rho$. This $\rho$ is the ratio of the total time a satellite spends in sight of its ground station to total flight time. Specifically\\
\begin{equation}
\label{RhoLimFormula1}
\rho=\lim\limits_{T\to\infty}\dfrac{P(T)}{T}
\end{equation}
Where, $T$ is the total flight time, and $P(T)$ is the time spent with line of sight between the satellite and the ground station. So, for a given flight time $T$, the total line of sight visibility time should be approximated by:\\
\begin{equation}
\label{RhoApproxFormula}
V(T)=\rho T
\end{equation}
Thus, given that the limit exists we would expect that as $T$ grows, $V(T)$ should become a better approximation for $P(T)$. That is,\\
\begin{equation}
\label{RhoLimFormula2}
\lim\limits_{T\to\infty}|V(T)-P(T)|=0
\end{equation}
Thus, for sufficiently long time periods, we may obtain a good approximation of $P(T)$ using $\rho$ alone.\\
For example, for an elliptical orbit with $a=10,000.14 \text{km},\;e=.2,\;i=28.5\degree,\;\lambda=0\degree$, (where $a=$ the major axis, $e=$ the eccentricity, $i=$ the inclination of the orbit and, $\lambda=$ the latitude of the ground station), we have:\\
$$P(1000\text{ days})=258.7599\text{ days}$$
$$V(1000\text{ days})=258.7937\text{ days}$$
$$\text{Percent Error}=.0131\%$$
Where $P(T)$ was computed by direct orbit propagation, then summing over all view periods in that time interval. $V(T)$ was computed as in Eq. \eqref{RhoApproxFormula}. $\rho$ was computed as the ratio: $\dfrac{P(T)}{T}$, by direct orbit propagation for $T=6000$ days (or, approximately 16 years), beyond which the changes to $\rho$ for larger $T$ became negligible.\\

\section[The Linear $J_2$ Model of Orbits]{The Linear $\boldsymbol{J_2}$ Model of Orbits}
As a result of the rotation of the Earth about its axis, it is not a perfect sphere, but an oblate spheroid. Because of this, the equations of motion governing the motion of an object orbiting it are perturbed. This perturbation can be quantified through the use of spherical harmonics, yielding a sequence of coefficients $J_2,J_3,...$ the normalized zonal harmonic gravitational coefficients. $J_2$ is by far the most significant of the coefficients, with $J_2=1.083\cdot10^{-3}$ and $J_3=-2.5\cdot10^{-6}$ (these are the values for Earth).

\subsection[The Nonlinear $J_2$ Model]{The Nonlinear $\boldsymbol{J_2}$ Model}
Neglecting higher order terms, under the $J_2$ perturbation the equations of motion which govern the motion of a body of negligible mass orbiting a planet are as follows\cite{Bate}\\
\begin{equation}
\begin{aligned}
\label{NonlinMotionEqs}
\ddot{x}&=-\dfrac{\mu x}{r^3}(1-J_2\dfrac{3R_B^2}{2r^2}(\dfrac{5z^2}{r^2}-1))\\
\ddot{y}&=\dfrac{y}{x}\ddot{x}\\
\ddot{z}&=-\dfrac{\mu z}{r^3}(1+J_2\dfrac{3R_B^3}{2r^3}(3-\dfrac{5x^2}{r^2}))\\
r&=(x^2+y^2+z^2)^\frac{1}{2}
\end{aligned}
\end{equation}
(Where $R_B$ and $\mu$ are the equatorial radius and gravitational parameter of the body respectively)\\
Note: This is in the non-inertial, body fixed rotating frame (rotating at the rate of the body's rotation about its axis).

\subsection[The Linear $J_2$ Model]{The Linear $\boldsymbol{J_2}$ Model}
A good approximation to these equations of motion may be found if we first pass into the state-space, $\Gamma$, of orbital elements such that\\
\begin{equation}
\begin{aligned}
\Gamma &= M\times\Omega\times\omega\text{, where}\\
M &= \text{Mean Anomaly,}\\
\Omega &= \text{Longitude of the Ascending Node,}\\
\omega &= \text{Argument of the Periapsis.}\\
\label{OrbEls}
\end{aligned}
\end{equation}
Then, linearizing these elements with respect to time, we get the following formulae approximating their time evolution \cite{Bate}\\
\begin{equation}
\begin{aligned}
\label{LinOrbEls}
M(t)&=M_0+\dot{M}t\\
\Omega(t)&=\Omega_0+\dot{\Omega}t \text{, mod }2\pi\\
\omega(t)&=\omega_0+\dot{\omega}t \text{, mod }2\pi
\end{aligned}
\end{equation}
Where their time derivatives are given by\\
\begin{equation}
\begin{aligned}
\label{LinOrbElConsts}
\dot{M} &= \sqrt{\dfrac{\mu}{a^3}}[1+3J_2\dfrac{R_B^2}{4a(1-e^2)^{\frac{3}{2}}}(3\cos^2(i)-1)]\\
\dot{\Omega} &= -\dfrac{3\sqrt{\mu}J_2R_B^2\cos(i)}{2a^{\frac{7}{2}}(1-e^2)^2}+\Omega_B\\
\dot{\omega} &= \dfrac{3\sqrt{\mu}J_2R_B^2(4-5\sin^2(i))}{4a^{\frac{7}{2}}(1-e^2)^2}
\end{aligned}
\end{equation}

\section{The Ergodic Theory of Satellite Coverage}
A dynamical system is considered to be ergodic if its trajectory repeatedly visits almost everywhere (in the measure-theoretic sense) in its state space \cite{Arnold}. The Birkhoff-Khinchin Theorem states that if a a system is ergodic then for every $f\in\mathcal{L}^1$ (where $\mathcal{L}^1$ is the set of functions which are absolutely Lebesgue integrable) defined on the state space of that system, the time mean of $f$ is approaches to the space mean of $f$ as the run time of the system grows.\cite{Arnold}\; \cite{Sinai} That is, if $\vec{x}(t)$ is the state of a system, $S$, at a time $t$, then if $f\in\mathcal{L}^1$:\\
\begin{equation}
\label{TimeSpace}
\lim_{T\to\infty}\dfrac{1}{T}\int\limits_{t=0}^Tf(\vec{x}(t))dt=\int\limits_{S}fd\mu
\end{equation}

In the case of satellite trajectories, we're interested in computing the time mean for the satellite in the region of space in which the satellite has line of sight communication with its ground station. More specifically, we want to compute the mean time that the satellite spends in the intersection of the region of visibility of the ground station, $V$, with the spatial state space, $S$, of the satellite, where $S$ is defined as the 3-dimensional region of space that the satellite can travel. Note that $V\cap S$ is the only region in which the satellite and ground station can feasibly communicate. This is because in the volume outside $V$, the line of sight communication isn't possible; and outside of $S$, the satellite cannot travel. We define this volume, $V^*=V\cap S$ as the region of feasible visibility.\\

\begin{figure}[htb]
\centering
\subfigure[Spatial State Space of Satellite in Elliptical Orbit]{
\centering
\includegraphics[width=.31\textwidth]{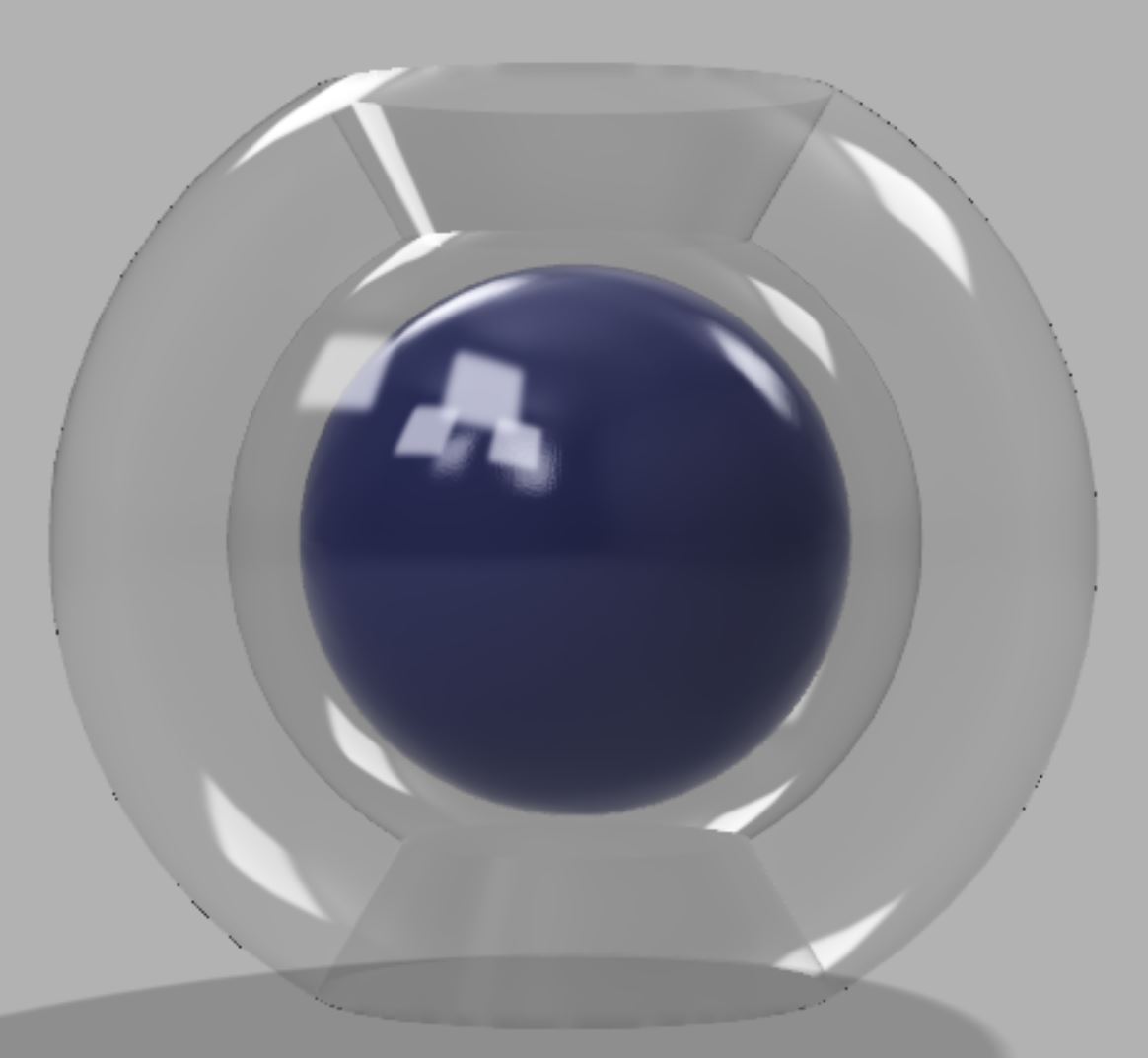}
\label{fig:1a}
}
\subfigure[Cross Section of Spatial State-Space]{
\centering
\includegraphics[width=.31\textwidth]{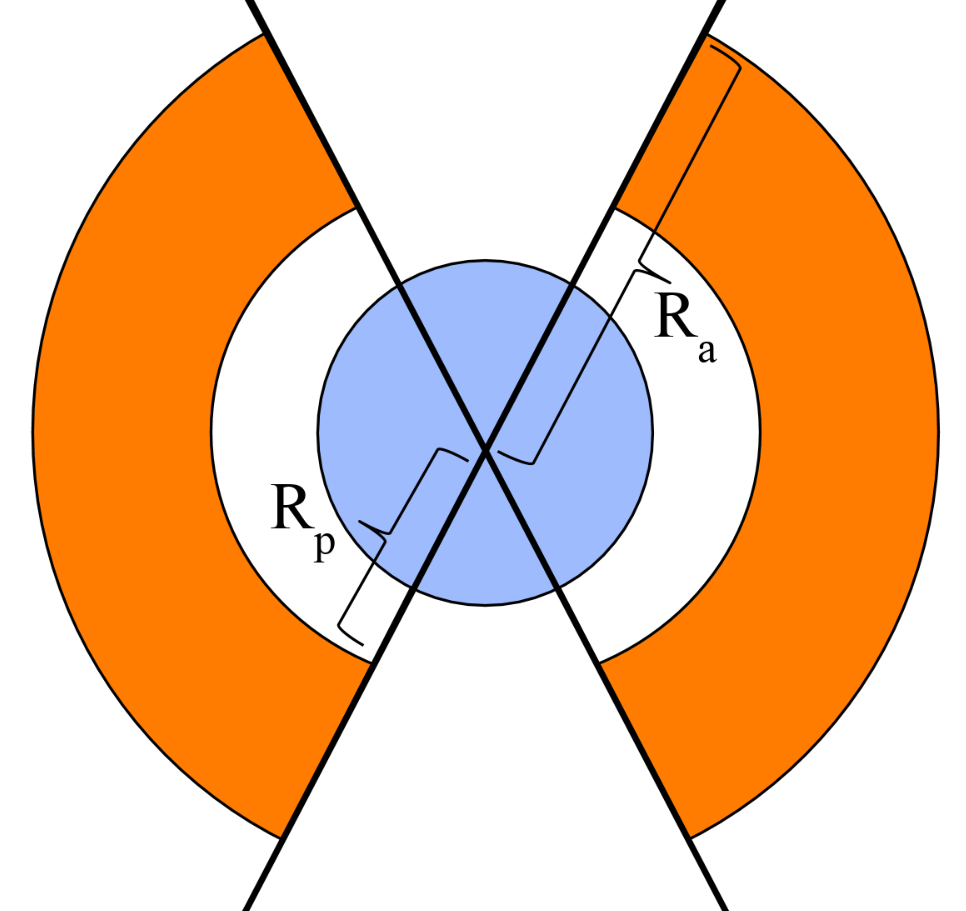}
\label{fig:1b}
}
\subfigure[Intersection of Spatial State Space with Ground Station Visibility Region]{
\centering
\includegraphics[width=.31\textwidth]{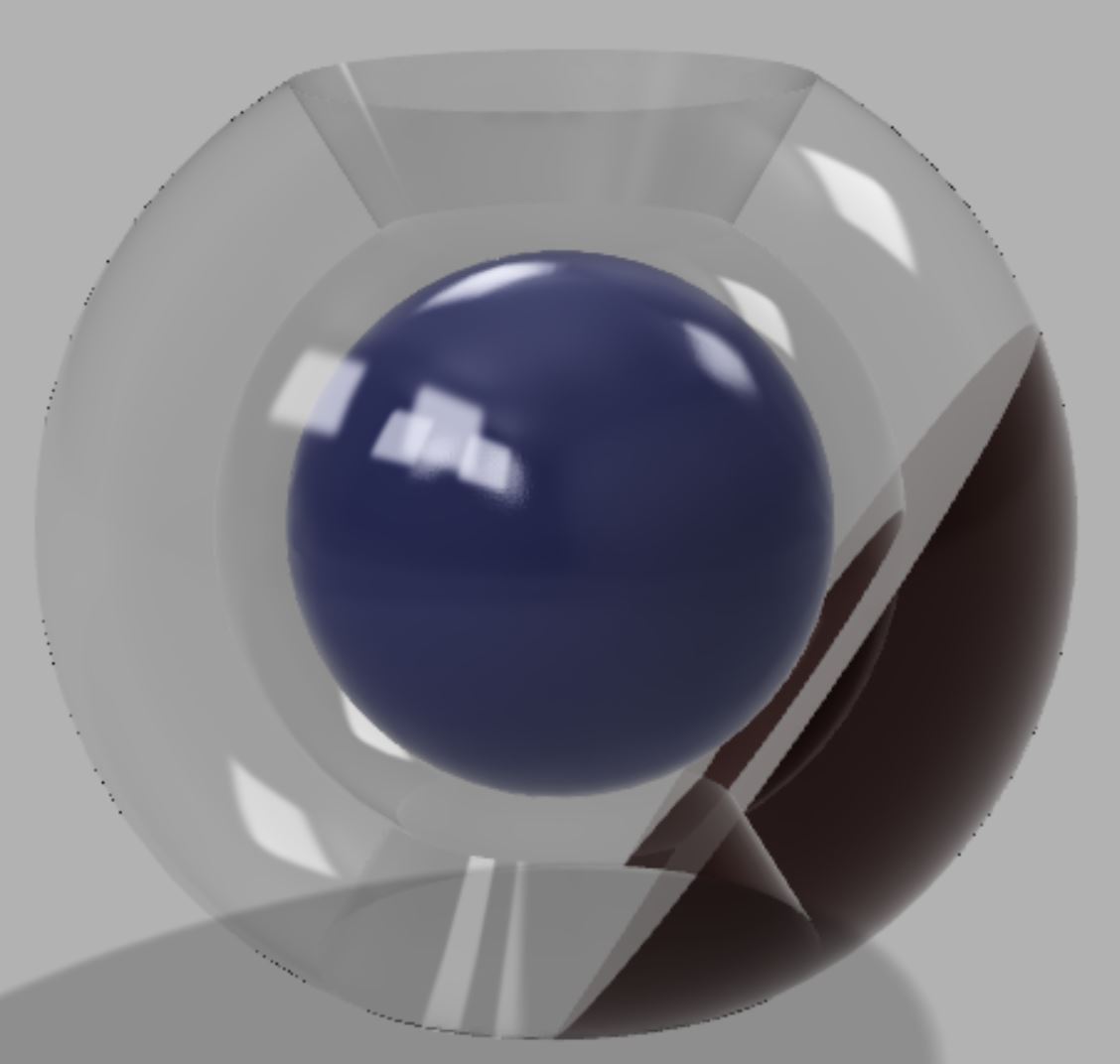}
\label{fig:1c}
}
\caption{Spatial State Space Visualizations}
\label{fig:1}
\end{figure}
\FloatBarrier 
Figure \ref{fig:1a} provides the visualization of the geometry of $S$ for a satellite in an elliptical orbit.
Figure \ref{fig:1b} shows a 2D cross section of the spatial state space, where $R_p$ is the periapsis and $R_a$ is the apoapsis.
Figure \ref{fig:1c} shows $V^*=V\cap S$ (in red) for a low latitude ground station. Note that for circular orbits, $S$ and $V^*$ reduce to 2-dimensional surfaces.\\

Now, applying the Birkhoff-Khinchin Theorem, we should expect a relationship between the ratio: $\frac{\text{Vol}(V^*)}{\text{Vol}(S)}$ and the mean time spent in that region. However, there are complications resulting from the fact that the mean density of the position of the satellite in this spatial state space is not uniform. More specifically, the satellite spends a disproportionately large amount of its flight time at extreme latitudes and near the apoapsis of its trajectory. Thus, there isn't a direct correspondence between the volume of a region of the satellite's state space and the time it spends there. That is, the spatial state space of the satellite is not ergodic. The bias of more extreme latitudes is shown below with the space tracks of a circular orbit:\\
\begin{figure}[htb]
\centering
\includegraphics[width=.45\textwidth]{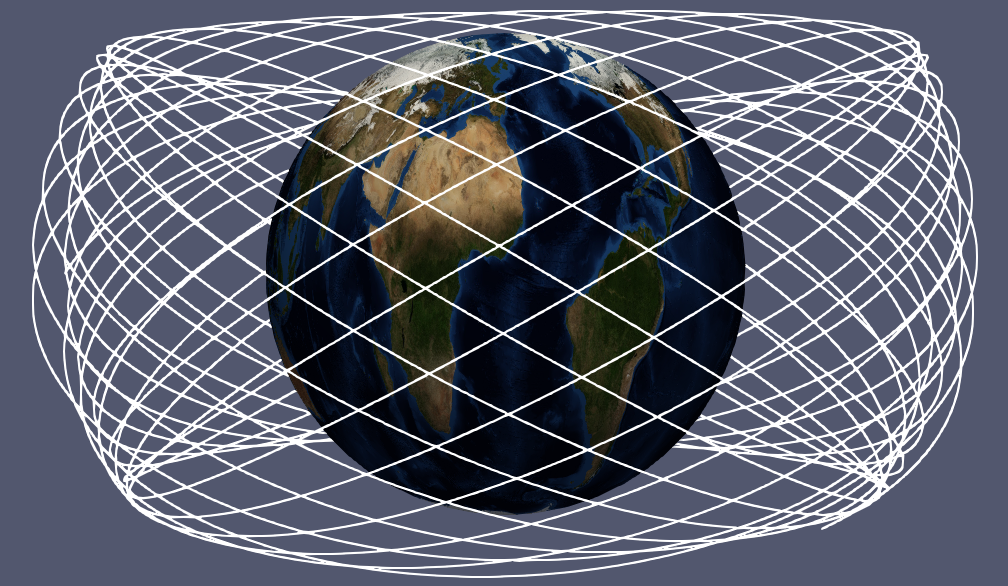}
\caption{Space Tracks of Satellite in Circular Orbit}
\label{fig:4}
\end{figure}
\FloatBarrier 
However, the linearized orbital element state space, $\Gamma$, of the satellite defined in Eq. \eqref{OrbEls}  is ergodic. This is because the flow (trajectory) is linear in time, hence the trajectory spends equal time in each region of $\Gamma$. Thus, $\frac{\text{Vol}(V^*_\Gamma)}{\text{Vol}(\Gamma)}$ should approximate the ratio of time the satellite spends in $V^*$. Where $V^*_\Gamma$ is the region of visibility for the ground station in $\Gamma$.

\section{The View Period Ratio Integrals}
In each of the cases below, we compute $\rho=\frac{\text{Vol}(V^*_\Gamma)}{\text{Vol}(\Gamma)}$. However, the geometry of the region of integration in the space of orbital elements, $\Gamma$, is sufficiently complicated that direct integration in this space is difficult. To resolve this, we can pull the integral back into the original space, $S$, where the geometry is much simpler. Let $\Phi:S\rightarrow\Gamma$ be a change of variables from $\Gamma$ to $S$, then the volume of the region in Gamma can be computed as an integral in S with the volume element $|\text{Det}[D\Phi]|$. This yields the desired formulae below.\\
\raggedbottom
\subsection{View Period Ratio for Circular Orbits}
\begin{equation}
\begin{aligned}
\rho(i,\lambda_s) &= \int\limits_{V^*}\dfrac{\cos(\lambda)}{2\pi^2\sqrt{\sin^2(i)-\sin^2(\lambda)}}dLd\lambda\\
\lambda_s &= \text{Latitude of Ground Station (Degrees)}\\
i &= \text{Inclination of Orbital Plane (Radians)}\\
\lambda &= \text{Latitude}\\
L &= \text{Longitude (Radians)}
\end{aligned}
\end{equation}

\subsection{View Period Ratio for Elliptical Orbits}
\begin{equation}
\begin{aligned}
\label{EllipticalIntegral}
\rho(a,e,i,\lambda_s) &= \dfrac{1}{2\pi^3}\int_{V^*}\dfrac{r\cos(\lambda )}{a\sqrt{\sin^2(i)-\sin^2(\lambda )}\sqrt{a^2e^2-(a-r)^2}}dLd\lambda dr\\
a &= \text{Major Axis (Kilometers)}\\
e &= \text{Eccentricity}\\
i &= \text{Inclination of Orbital Plane (Degrees)}\\
\lambda_s &= \text{Latitude of Ground Station (Degrees)}\\
r &= \text{Radius from Center of Body (km)}\\
\lambda &= \text{Latitude (Radians)}\\
L &= \text{Longitude (Radians)}
\end{aligned}
\end{equation}

\section{Numerical Verification}
\subsection{Evaluating the Integral}
In this section, we restrict our focus to the numerical evaluation of $\rho$ in the elliptical case. A fast technique for computing the integral for circular orbits is given by Lo\cite{Lo1994}. Furthermore, the elliptical formula may be used to compute the ratio for circular orbits by setting the eccentricity, $e<<1$.

As a result of the irregular shape of $V^*$ and multiple singularities in the integrand, accurately evaluating the definite integral over $V^*$ has proven to be a difficult problem. Several different potential techniques for evaluating the integral are discussed below.\\
\subsubsection{Monte Carlo:}
Due to it's speed and ease of implementation, early on Monte Carlo Integration was method of choice for evaluating the integral. For early verification of the formula's accuracy it was a great tool, producing decently accurate results. Unfortunately, by the nature of Monte Carlo, it is difficult to get beyond two to three or so decimal places of accuracy. Which is problematic if the ratio being calculated begins in the second decimal place, as many did. It is possible to increase accuracy by increasing the number of trials but this comes at the cost of the calculation becoming impractically long. Monte Carlo was eventually abandoned as a method of evaluation for these reasons.
\subsubsection{Direct Integration:}
Another technique which was investigated is direct definite integration over the region using casework, with a different case for each distinct shape of the region. The shape of the region can be affected by many factors, such as whether it intersects the edges of the state space, and the shape of the region of visibility. Initial results for this technique were good. Where the casework method worked, it worked well, yielding five to six decimal places of accuracy in less than a second. However, there were many cases for which the region was simply too complex to admit a comprehensible definite integral form (such as for nearly polar ground stations). In addition, this technique wouldn't be very portable to other geometries of ground station visibility regions due to the amount of case work required.
\subsubsection{Direct Integration with Indicator Function:}
Another option is integration over the entire state space with an indicator function on the region of interest. This was discovered to be by far the most successful of the three methods. It retained the versatility of the Monte Carlo method, being capable of effectively integrating over even quite pathological volumes. While also maintaining the level of accuracy offered by the casework definite integration method. This was at the cost, however, of speed. Although much faster than Monte Carlo, integrating over such a large region when only a small portion of it is actually non-zero forces the numerical integrator to compute a very large number of function evaluations. Below, a brief explanation/derivation of the indicator function method is given. Taking Eq. \eqref{EllipticalIntegral}:\\
\begin{equation*}
\begin{aligned}
\rho(a,e,i,\lambda_s) = \dfrac{1}{2\pi^3}\int_{V^*}\dfrac{r\cos(\lambda )}{a\sqrt{\sin^2(i)-\sin^2(\lambda )}\sqrt{a^2e^2-(a-r)^2}}dLd\lambda dr
\end{aligned}
\end{equation*}
The idea is to integrate over the entire spatial state space, defined in part 2, multiplying the integrand by the indicator function, $\mathbf{1}_V$, on the region of visibility of the ground station, $V$:\\
\begin{equation}
\begin{aligned}
\rho(a,e,i,\lambda_s) = \dfrac{1}{2\pi^3}\int_S \dfrac{r\cos(\lambda)\mathbf{1}_V(r,\lambda,L)}{a\sqrt{\sin^2(i)-\sin^2(\lambda )}\sqrt{a^2e^2-(a-r)^2}}dLd\lambda dr
\end{aligned}
\end{equation}
Which becomes:\\
\begin{equation}
\begin{aligned}
\rho(a,e,i,\lambda_s) = \dfrac{1}{2\pi^3}\int\limits_{r=a(1-e)}^{a(1+e)}\int\limits_{\lambda=-i}^i\int\limits_{L=0}^{2\pi} \dfrac{r\cos(\lambda)\mathbf{1}_V(r,\lambda,L)}{a\sqrt{\sin^2(i)-\sin^2(\lambda )}\sqrt{a^2e^2-(a-r)^2}}dLd\lambda dr
\end{aligned}
\end{equation}
It was discovered that applying the following change of variables simplifies the integral considerably:\\
\begin{equation}
\begin{aligned}
\label{ChangeofVariables}
r(\theta) &=a(1-e\sin(\theta))\\
\lambda(\alpha) &= \sin^{-1}(\sin(i)\sin(\alpha))
\end{aligned}
\end{equation}
We get:\\
\begin{equation}
\begin{aligned}
\label{SimplifiedEllipticalIntegral}
\rho(a,e,i,\lambda_s) = \dfrac{1}{2\pi^3}\int\limits_{\theta=-\frac{\pi}{2}}^{\frac{\pi}{2}}\int\limits_{\alpha=-\frac{\pi}{2}}^{\frac{\pi}{2}}\int\limits_{L=0}^{2\pi}\mathbf{1}_V(r(\theta),\lambda(\alpha),L)(1-e\sin(\theta))dLd\alpha d\theta
\end{aligned}
\end{equation}

This integral clearly contains far fewer operations per function evaluation. In addition, this form of the formula has no singularities, which reduces integration time significantly.
\subsection{Generating the Test Data}
For the purposes of testing the formula, the elevation angle of the line of sight for the ground station was set to zero. That is, the satellite was assumed to be able to communicate with the ground station if it was above the plane tangent to the planet at the location of the ground station.

After selecting the set of points to test (in next section), a method of comparing the results generated via Eq. \eqref{EllipticalIntegral} to the real world (or some good approximation of the real world) was needed. Generating the formula data was relatively fast, only taking a couple of hours to sequentially compute ratios for 6000 data points. The method of choice for generating the "real world" data was the classical method mentioned above, integrating the trajectory of the satellite with a numerical ordinary differential equation solver.

However, this approach proves problematic because of how slow it is. Using the Matlab implementation of ode45, run for 6000 simulation days, it conservatively took an hour for the computation to complete. Which would amount to computing 8000 data points taking on the order of a year to complete ($\sim$333 days), which is not an acceptable run time. Luckily, another research group in the same department was working on parallel numerical ode integration using GPUs. The GPU group was able to produce an integrator which would work well for integrating these satellite trajectories. This brought down the compute time down from a year to several hours.
\subsection{Verification of Results}
Figure \ref{fig:5} below shows the residual error between the Ergodic results and the integrated (direct orbit propagation) values for Earth. Each data point was randomly sampled from the below sample space (Monte Carlo):\\
\begin{equation*}
\begin{aligned}
\label{SampleSpace1}
100+R<&a<50000+R\\
.001<&e<.95\\
0<&i<\pi\\
-\dfrac{\pi}{2}<&\lambda<\dfrac{\pi}{2}\\
R &= \text{Radius of planet (km)}\\
a &= \text{Major Axis (km)}\\
e &= \text{Eccentricity}\\
i &= \text{Inclination angle of orbital plane (radians)}\\
\lambda &= \text{Latitude of ground station (radians)}
\end{aligned}
\end{equation*}
With the restriction on each data point that:\\
1. The eccentricity not be so large that the trajectory would intersect the planet.\\
2. That the ground station latitude is no larger than the inclination angle.\\
3. The trajectory is non-periodic.\\
\begin{figure}[hptb]
\centering
\includegraphics[width=.5\textwidth]{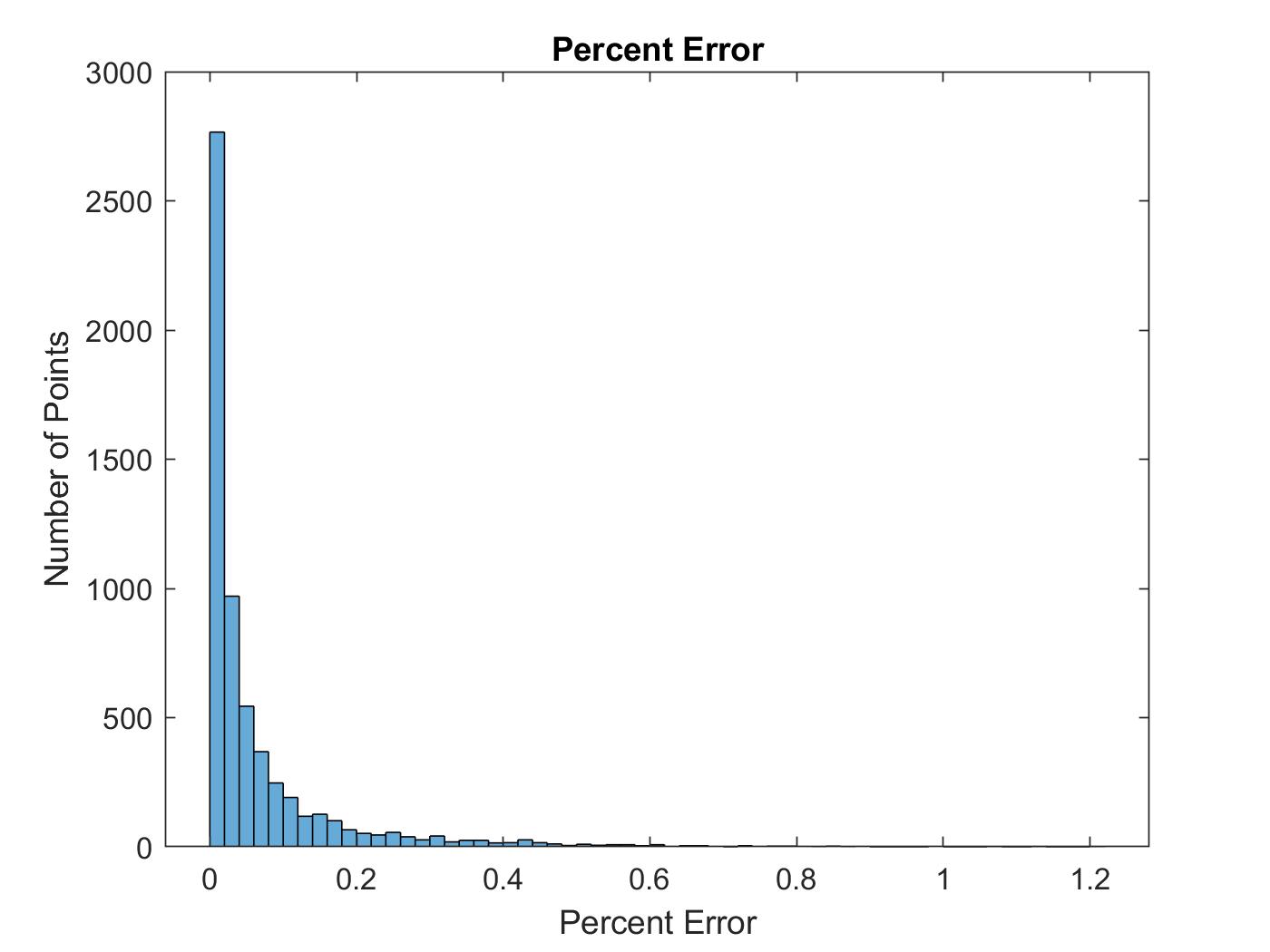}
\caption{Histogram of Percent Error}
\label{fig:5}
\end{figure}
\FloatBarrier
\begin{tabular}{l l l}
&Max percent error &=\;\;\; $1.22$\\
&Min percent error &=\;\;\; $8.38\times 10^{-8}$\\
&Mean percent error &=\;\;\; $.0658$\\
&Median percent error &=\;\;\; $.0239$
\end{tabular}

\noindent Next, a more systematic method of sampling was used, with each feasible point in the below range of values for the inputs included.\\
\begin{tabular}{l l l}
&Altitude &$\in \{100,200,300,400,500,1000,2000,5000,10000,25000,50000\}$ (kilometers)\\
&Eccentricity &$\in \{.01,.05,.1,.2,.3,.4,.5,.6,.7,.8,.9\}$\\
&Inclination &$\in \{5,10,15,...,80,85\}$ (degrees)\\
&GS Latitude &$\in \{0,5,10,...,85,90\}$ (degrees)
\end{tabular}
With the restriction on each data point that:\\
1. The eccentricity not be so large that the trajectory would intersect the planet.\\
2. The trajectory is non-periodic.\\
\begin{figure}[hptb]
\centering
\includegraphics[width=.5\textwidth]{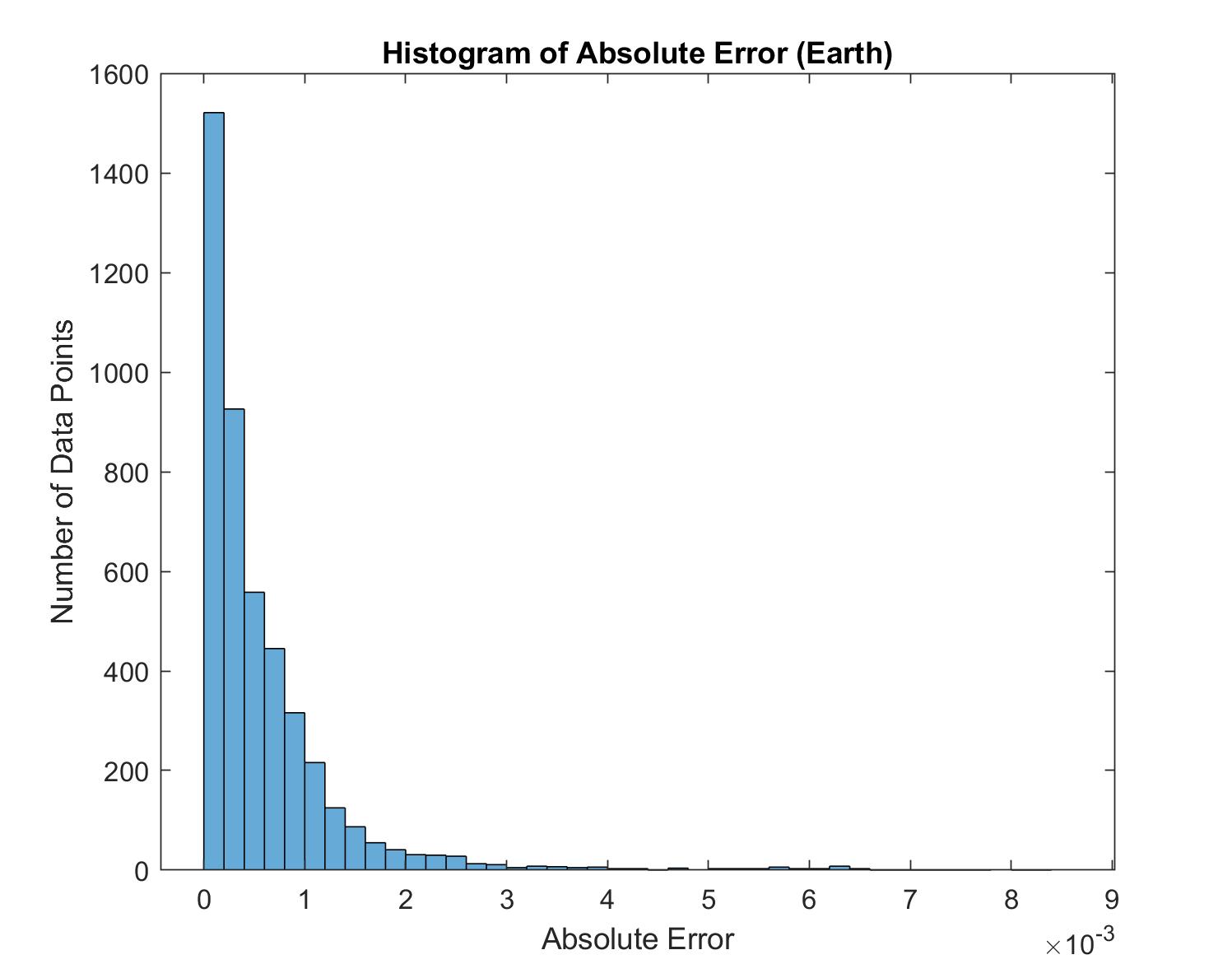}\\
\caption{Histogram of Absolute Error}
\label{fig:6}
\end{figure}\\
The above chart is a histogram of the absolute error between the ergodic and numerical results for Earth. It is everywhere less than .01, which implies that the ergodic result will be off by, at most 1 percentage point. Moreover, the mean absolute error is .00058. So, on average, the theoretical result will be off by .058 percentage points.
\begin{figure}[hpt]
\subfigure[Percent Error Unsorted]{
\centering
\includegraphics[width=.5\linewidth]{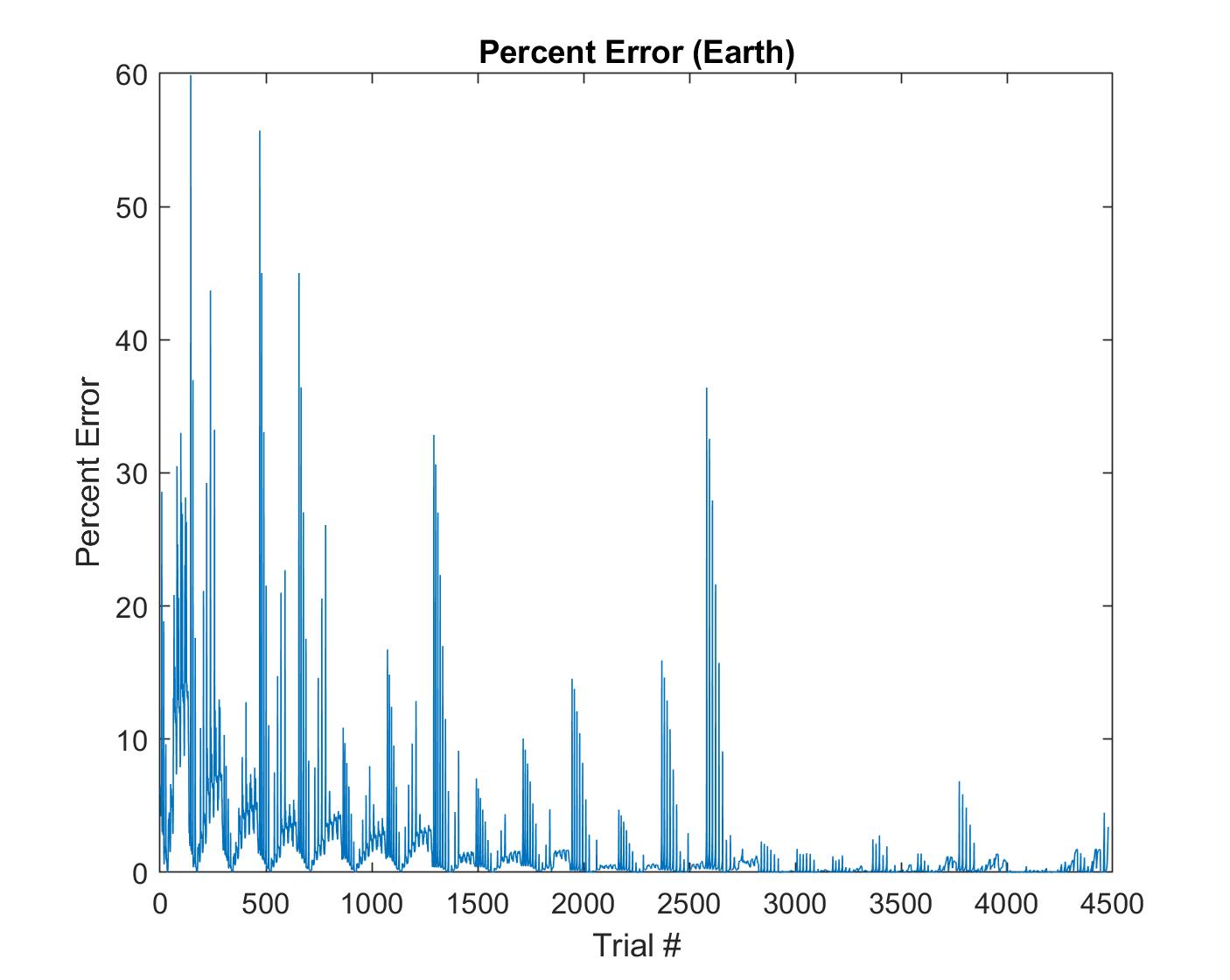}
\label{fig:7}
}
\subfigure[Percent Error Sorted by Ratio Size]{
\centering
\includegraphics[width=.5\linewidth]{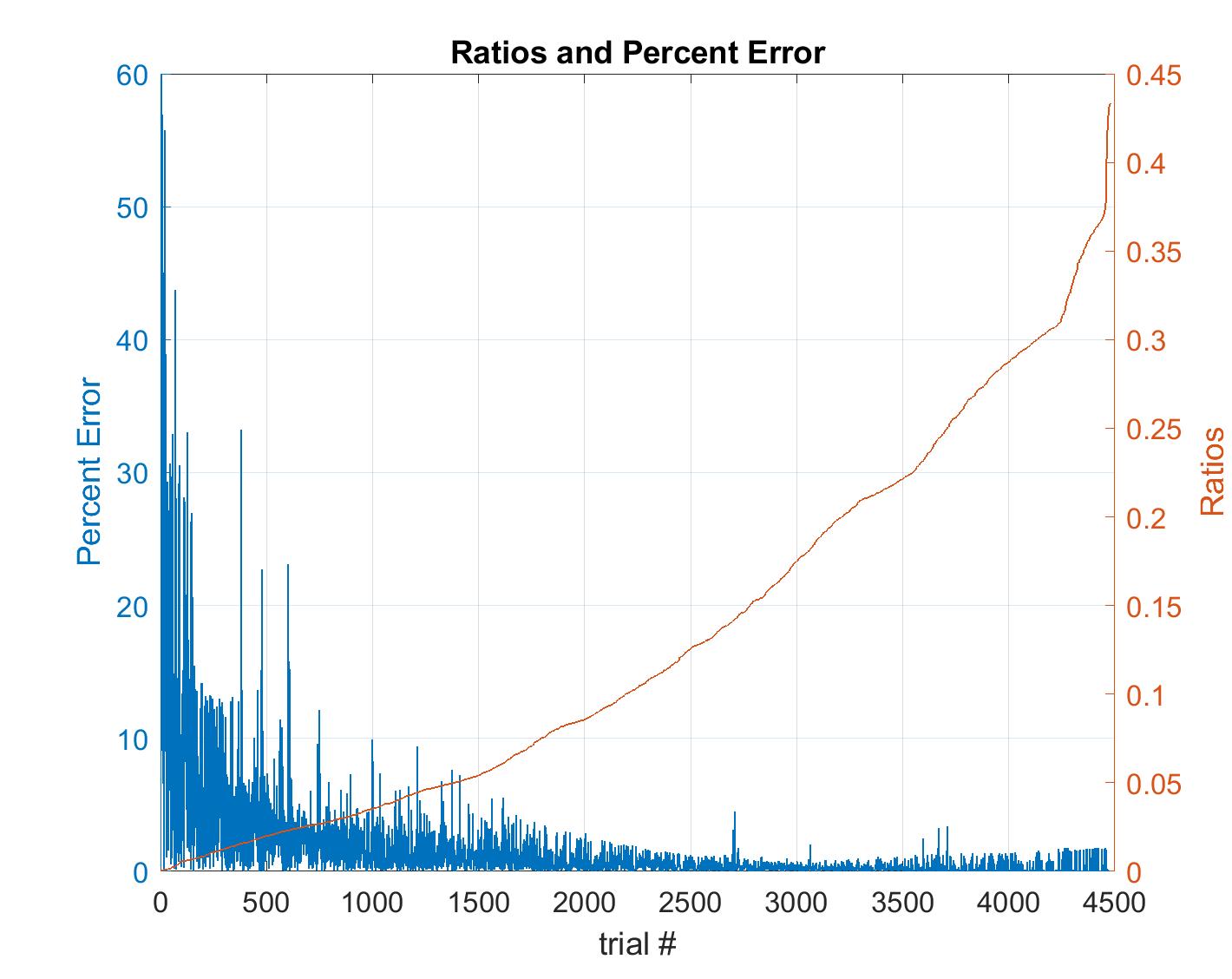}
\label{fig:8}
}
\caption{Percent Error}
\end{figure}
\FloatBarrier 
The percent error numbers on Fig. \ref{fig:7} make the situation look far worse. In places, percent error is as high as 60\%. This is a result of the fact that if the ratio is sufficiently small, even very small error ends up being a considerable percentage of that ratio. Fig. \ref{fig:8} is sorted by ratio size from smallest to largest (with the ratio magnitude in orange and percent error in blue). Clearly, the results with very large percent error are restricted to the smallest ratios, where percent error is expected to be larger.

\section{Conclusion}
\raggedbottom
The fact that very large portion of satellite trajectories are periodic would seem to impose a severe restriction on the usefulness of this formula. However, it has been shown that the formula still tends to give reasonably good results for periodic orbits. In addition, the exact result may be computed for any periodic orbit by propagating the trajectory for one full period, which is much cheaper than long-term orbit propagation. Another issue encountered was the fact that percent error blows up when the visibility ratio becomes sufficiently small. This may be somewhat problematic. However, trajectories where this sort of issue happens are relatively rare in practice. For example, low altitude trajectories (100-150 km) tend to lead to small ratios where this problem occurs. But such low Earth orbits are uncommon in practice due to atmospheric drag among other issues (the ISS stays at least 300 km above the Earth and still has to make relatively large corrections due to drag). In general, configurations with small ratios will often be of little interest in comparison to high ratio configurations where the percent error is small.

Another issue with the visibility ratio for elliptical orbits is that it requires a volume integral which can be very slow to compute; whereas for circular orbits, the visibility ratio is given by a simple definite integral that can be quickly integrated numerically. To solve this problem, we plan to implement a GPU based algorithm to compute the volume integral. We expect this to reduce the computation time by several orders of magnitude. By implementing a GPU-based function to compute the visibility ratio of elliptical orbits such as in Matlab or Python, this would provide engineers and designers with a fast and powerful tool for designing telecommunications systems and network in real time.
\section{Acknowledgments}
This research was carried out in part at the Jet Propulsion Laboratory, California Institute of Technology, and was sponsored by Caltech Summer Undergraduate Research Fellowship Program and the National Aeronautics and Space Administration. This work was also supported in part by the Hummer-Tuttle gift to Professor Al Barr through the Caltech Division of Engineering and Applied Science. Thanks to Ryan Burns, Brian Anderson, Max Zhan, Tom Gorordo and Sam Blazes for assistance with GPU programming, data generation, and visualization.

\section{Notation}

{\renewcommand\arraystretch{1.0}
\noindent\begin{longtable}{@{}l @{\quad=\quad} l@{}}
$\rho$ & Satellite-Ground Station Visibility Ratio\\
$T$ & Total Flight Time\\
$P(T)$ & Total Line of Sight Time over $T$\\
$V(T)$ & Expected Line of Sight Time over $T$ (given $\rho$)\\ 
$a$ & Major Axis (Kilometers)\\
$e$ & Eccentricity\\
$i$ & Inclination of Orbital Plane (Degrees)\\
$\lambda_s$ & Latitude of Ground Station (Degrees)\\
$L$ & Longitude (Degrees)\\
$R_B$ & Radius of Body (Kilometers)\\
$\mu$ & The Standard Gravitational Parameter of the Body ($\text{km}^3\text{s}^{-2}$)\\
$M$ & Mean Anomaly (Radians)\\
$\Omega$ & Longitude of the Ascending Node (Radians)\\
$\omega$ & Argument of the Periapsis (Radians)\\
$J_2$ & The Body’s Oblateness Parameter\\
$\mathcal{L}^1$ & $\{f|(\int_{S}|f|d\mu)<\infty\}$\\
$S$ & Spatial State Space of Satellite\\
$V$ & Spatial Region of Visibility for Ground Station\\
$V^*$ & The Region of Feasible Visibility\\
$\Gamma$ & The Orbital Elements State Space of the Satellite\\
$V_\Gamma$ & Region of Visibility for Ground Station in $\Gamma$\\
$V^*_\Gamma$ &  The Region of Feasible Visibility in $\Gamma$
\end{longtable}}

\bibliographystyle{AAS_publication}   
\bibliography{references}   

\end{document}